# Experimental determination of the topological phase diagram in Cerium monopnictides

Kenta Kuroda,[1] M. Ochi,[2] H. S. Suzuki,[1,3] M. Hirayama,[4] M. Nakayama,[1] R. Noguchi,[1]
C. Bareille,[1] S. Akebi,[1] S. Kunisada,[1] T. Muro,[5] M. D. Watson,[6] H. Kitazawa,[3]
Y. Haga,[7] T. K. Kim,[6] M. Hoesch,[6] S. Shin,[1] R. Arita,[4] and Takeshi Kondo[1]

[1]*ISSP, University of Tokyo, Kashiwa, Chiba 277-8581, Japan*
[2]*Department of Physics, Osaka University, Machikaneyama-cho, Toyonaka, Osaka 560-0043, Japan*
[3]*National Institute for Materials Science, 1-2-1 Sengen, Tsukuba 305-0047, Japan*
[4]*RIKEN Center for Emergent Matter Science (CEMS), 2-1 Hirosawa, Wako, Saitama 351-0198, Japan*
[5]*Japan Synchrotron Radiation Research Institute (JASRI), 1-1-1 Kouto, Sayo, Hyogo 679-5198, Japan*
[6]*Diamond Light Source, Harwell Campus, Didcot, OX11 0DE, United Kingdom*
[7]*Advanced Science Research Center, Japan Atomic Energy Agency, Tokai, Ibaraki 319-1195, Japan*
(Dated: July 20, 2017)

We use bulk-sensitive soft X-ray angle-resolved photoemission spectroscopy and investigate bulk electronic structures of Ce monopnictides (CeX; X=P, As, Sb and Bi). By exploiting a paradigmatic study of the band structures as a function of their spin-orbit coupling (SOC), we draw the topological phase diagram of CeX and unambiguously reveal the topological phase transition from a trivial to a nontrivial regime in going from CeP to CeBi induced by the band inversion. The underlying mechanism of the topological phase transition is elucidated in terms of SOC in concert with their semimetallic band structures. Our comprehensive observations provide a new insight into the band topology hidden in the bulk of solid states.

The discovery of topological insulators represents a significant progress of topological band theory. [1–3]. They are characterized by nontrivial $Z_2$ topological invariant obtained if the conduction and valence bands with different parity are inverted due to spin-orbit coupling (SOC) [4]. In three-dimensional (3D) case, the band inversion gives a rise to topological surface states (TSSs) inside the energy gap. The concept was generalized to various systems, which has revealed great richness of intriguing topological phase such as Weyl semimetal [5–7] and Dirac nodal-line semimetal [8–10].

Owing to the bulk-edge correspondence [11, 12], investigations of the in-gap TSS in principle can indirectly display the band topology hidden in the bulk states. In fact, surface-sensitive angle-resolved photoemission spectroscopy with vacuum ultraviolet (VUV-ARPES) has achieved great success to confirm the Dirac-like TSS in a number of chalcogenides [13–17], which have obtained excellent agreements with the predicted nontrivial $Z_2$ topology [18–20].

However, searches of the topological phase are still challenging in low carrier semimetallic rare-earth monopnictides, with the NaCl-type crystal structure, LnX (Ln=La or Ce; X=P, As, Sb or Bi) [21–28]. The main difficulty comes from two issues. One is that first-principles calculation shows controversial conclusions about their band topology, since it often misestimates the band gap [21–23], and therefore the experimental determination is necessary. Secondly, despite of this, so far the experimental confirmations of the band topology are restricted only to the indirect investigation via the surface dispersions predicted by the calculations [23–27]. LaBi has been considered to be a topologically nontrivial state [24–27]. In LaSb, the Dirac-cone-like energy dispersion has been observed and however the interpretation is fully controversial: Zeng *et al.* reported that it originates from trivial bulk states [23] while Niu *et al.* concluded that it comes from the TSS [25]. The recent VUV-ARPES showed the Dirac-cone-like dispersion also in CeSb and CeBi [28] but the interpretation is unclear. Thus far, the unclarity of this indirect measurement with surface-sensitive probe poses the difficulty to elucidate the bulk band topology.

In this Letter, we present an alternative way to directly clarify the band topology by using bulk-sensitive soft X-ray ARPES (SX-ARPES) on CeX series materials. By this paradigmatic investigation of their electronic structures from CeP to CeBi, we draw the topological phase diagram of CeX as a function of their SOC. The obtained phase diagram unambiguously demonstrates the topological phase transition from a trivial to a nontrivial regime across the border between CeSb and CeBi.

Single crystalline CeX's was grown by using Bridgman method. Bulk-sensitive SX-ARPES measurement was performed at BL25SU at SPring-8 [29]. The photoelectrons were acquired by hemispherical analyzer ScientaOmicron DA30. The total experimental energy resolution was set to about 80 meV for photon energy ($h\nu$) of 500-760 eV. Surface-sensitive VUV-ARPES measurement was performed at I05 ARPES beamline at DIAMOND light source [30]. The total experimental energy resolution was set to below 20 meV for $h\nu$ of 25-100 eV. All samples were cleaved at a pressure of $5\times10^{-8}$ Pa at approximately 60 K, exposing shiny surfaces corresponding to the (001) plane. The sample temperature was kept at 60 K during the measurement to avoid magnetic

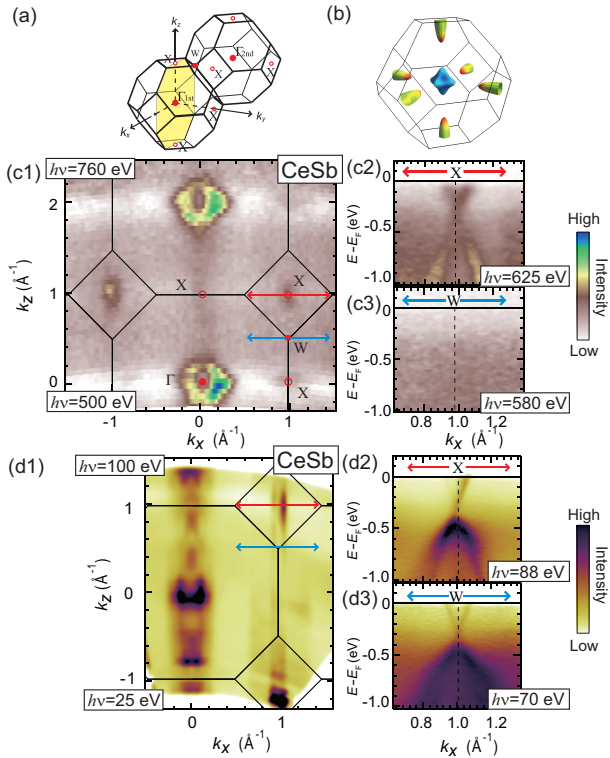

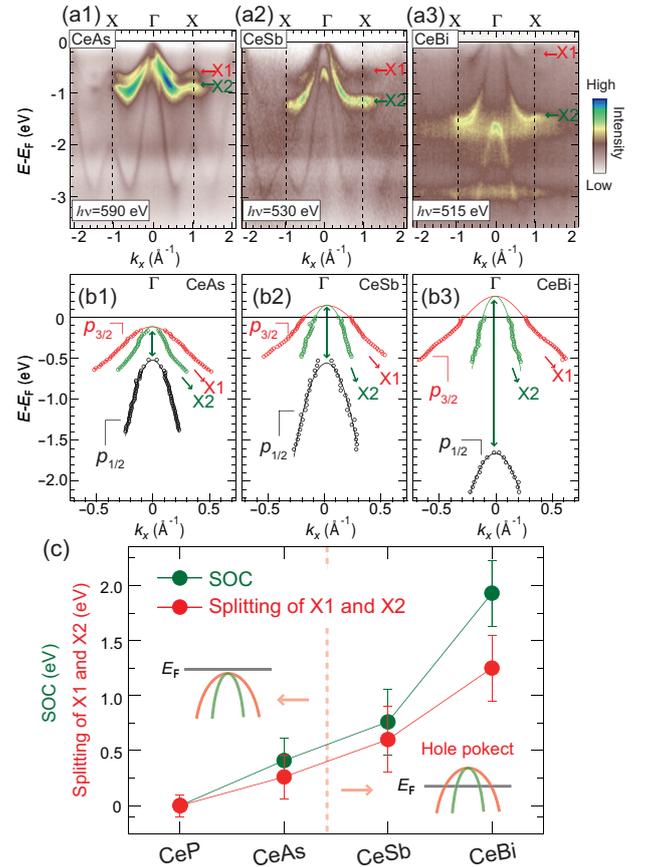

FIG. 1: (a) 3D Brillouin zone (BZ), showing the $k_z$-$k_x$ sheet at $k_y$=0. (b) Calculated bulk Fermi surfaces (FSs). (c1) SX-ARPES result for FS mapping on the $k_z$-$k_x$ plane [yellow plane in (a)] with different $h\nu$. (c2) and (c3) SX-ARPES band maps cut along different $k_x$ crossing X and W [red and blue lines in (c1), respectively]. (d1) VUV-ARPES FS mapping on the $k_z$-$k_x$ plane by changing $h\nu$. (d2) and (d3) VUV-ARPES band maps at different $k_z$ as shown in (d1).

FIG. 2: (a1)-(a3) SX-ARPES band maps for CeX's cut along X-Γ-X line. (b1)-(b3) Comparison of (red and green) the $p_{3/2}$ and (black) $p_{1/2}$ energy dispersions around Γ. The energy positions are determined by tracing the peak position of the momentum distribution curves. (c) Experimentally determined strength of their spin-orbit couplings. The size was defined as the energy difference for the top of $p_{3/2}$ and $p_{1/2}$ bands at Γ [green arrows in (b1)-(b3)].

phases at low temperature [31].

Electronic structure calculations were performed with the QUANTUM ESPRESSO package [32], using a relativistic version of pseudo potentials with Ce $f$-electrons treating as core electrons. We use the experimental crystal structure taken from Ref. [33]. For the construction of the Wannier orbital of Ce $t_{2g}$ and pnictogen $p$ orbitals, we employed the wannier90 code [34]. All calculation results presented in this paper were obtained using the tight-binding models constructed from the hopping parameters among the Wannier orbitals.

We start with presenting bulk-sensitive SX-ARPES and surface-sensitive VUV-ARPES results for Fermi surface (FS) mappings in CeSb. Figure 1(c1) presents the FS mapping on the $k_x$-$k_z$ plane [yellow plane in Fig. 1(a)] recorded with varying $h\nu$ from 500 eV to 760 eV. This data displays the clear $k_z$ dispersions and the FS topology of CeSb which is consistent with calculation in Fig. 1(b), showing an elliptical electron pocket formed by Ce $t_{2g}$ at X, and hole pockets originated from Sb 5$p$ at Γ.

In contrast, the $k_z$ dispersion is unclear in VUV-ARPES [Fig. 1(d1)]. Figures 1(d2) and (d3) show the VUV-ARPES maps along different $k_x$ cuts crossing X and W, respectively [red and blue lines in Fig. 1(d1)]. In the both $h\nu$, we observe the Dirac-cone-like energy dispersion, which is seen even by using different $h\nu$ in VUV range (Supplementary-Figure S1). This behavior is reminiscent of the surface state with no $k_z$-dependence. Indeed, the Dirac-cone dispersion in CeSb was previously interpreted as the in-gap TSS [28]. However, we here come to an alternative conclusion. SX-ARPES measurements under a precise $k_z$ definition [35] clearly observe the $k_z$ dependence for the same band [Figs. 1(c1)-(c3)]. These facts identify the Dirac-cone-like dispersion as the 3D bulk states. Apparently, VUV-ARPES loses the intrinsic band dispersion in CeSb. We therefore use SX-ARPES to investigate the band topology of CeX's.

We now turn to compare the bulk electronic structures of CeX's. By tuning $h\nu$ in SX range, we selectively ob-serve their bulk band dispersions in wide energy range along X-Γ-X in Figs. 2(a1)-(a3) (the data for CeP are presented in Supplementary-Figure S2). By systematically looking at their electronic structures, we find the SOC effect and its evolution with moving in the pnictogen from P to Bi. In Figs. 2(b1)-(b3), we show the detailed energy dispersions of the valence $p$ bands around Γ. Due to the SOC, the $p$ bands split into the $p_{1/2}$ and $p_{3/2}$ states. The splitting size is different for these compounds. Figure 2(c) represents the estimated size of their SOC (see the caption). In going from CeP to CeBi, the SOC strength becomes large from ∼0 eV up to ∼2 eV.

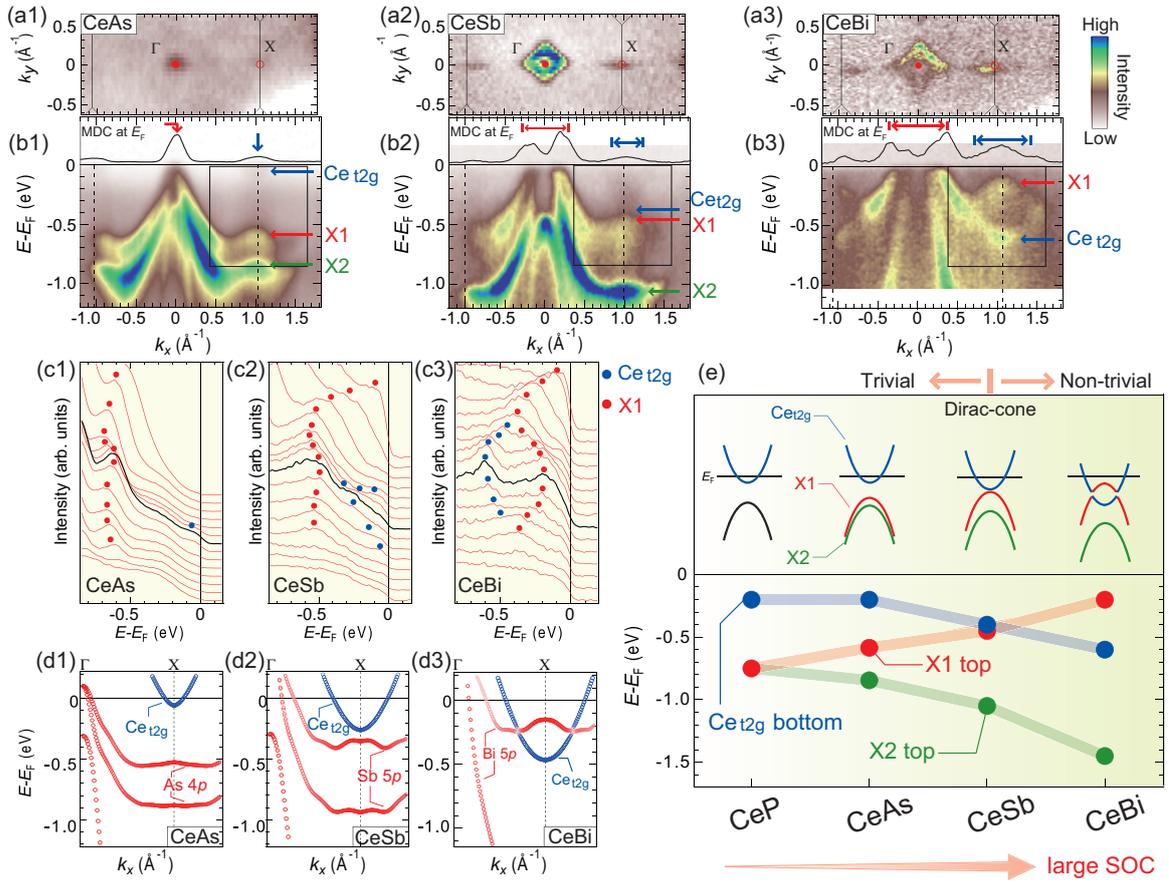

FIG. 3: (a1)-(a3) FS mapping on the $k_x$-$k_y$ plane at $k_z$=0 for CeX's. (b1)-(b3) Enlarged SX-ARPES band maps of Figs. 2(a1)-(a3) and (inset) their momentum distribution curves (MDCs) at $E_F$. The hole (electron) pockets are shown by red (blue) arrows above the MDCs. (c1)-(c3) The energy distribution curves (EDCs) around X within $E$-$k_x$ window in (b1)-(b3), displaying (red circles) X1 and (black circles) Ce $t_{2g}$ dispersions. The EDC at X is highlighted by the black line. (d1)-(d3) The calculated band structures around X with parity analysis. The red and blue represent the pnictogen $p$ and Ce $t_{2g}$ orbital contributions. For the calculation, we used the onsite-energy shift for Ce $t_{2g}$, $\Delta_{t2g}$=0.85 eV (CeAs), 0.60 eV (CeSb) and 0.54 eV (CeBi). (e) Experimentally determined topological phase diagram and (inset) the schematics of the band structures around X.

We find two important consequences of the SOC. First, the SOC induces the valence band splitting also at X [X1 and X2 bands shown in Figs. 2(a1)-(a3)]. The large SOC pushes X1 (X2) band up (down) in energy. The splitting becomes significant with increasing SOC [Fig. 2(c)]. Second, for CeSb and CeBi, the higher-lying $p_{3/2}$ bands are pushed above $E_F$ due to the large SOC, and the hole pockets appear [Figs. 2(b2) and (b3)]. Since the CeX series materials have the similar low carrier density [36], the SOC evolution of the hole pockets should increase the size of the Ce $t_{2g}$ electron pocket at X. This carrier compensation is necessary in the semimetallic structure of CeX.

The electron band evolution is captured in Figs. 3(a1)-(a3) and 3(b1)-(b3) where we present the FS mappings on the $k_x$-$k_y$ plane at $k_z$=0 and the enlarged band maps near $E_F$, respectively. The intensities from the Ce $t_{2g}$ band are seen in the MDCs at $E_F$ [insets of Figs. 3(b1)-(b3)]. For CeAs, the both of the hole and electron pockets are quite small [Figs. 3(a1) and (b1)]. With growing the hole pocket at Γ from CeAs to CeBi [Figs. 3(b1)-(b3)], the electron band bottom goes down in energy and the size of the electron pocket becomes large (blue arrows). In addition, one can trace X1 and X2 energy dis-persions in Figs. 3(b1)-(b3). They disperse downwards in energy from Γ to X, while turn to be weakly upward at X particularly in CeAs. The large splitting of X1 and X2 eventually leads to the band inversion in CeBi, which can be clearly seen in their EDCs around X as shown in Figs. 3(c1)-(c3). For CeAs, X1 (red circles) and Ce $t_{2g}$ (blue circles) are separated in energy. In going from CeAs to CeBi, the upward band dispersion of X1 and the downward dispersion of Ce $t_{2g}$ approach for CeSb forming the Dirac-cone-like energy dispersion with a small gap ∼0.1 eV, and these two bands are finally inverted in CeBi.

Figures 3(d1)-(d3) show the calculated band structure with analyzing the orbital parity for CeX. The band gap obtained by SX-ARPES measurements is available to determine the realistic band topology. To do this, we introduce the onsite-energy shift for Ce $t_{2g}$ orbitals in our tight-binding model (see the caption). The parity eigenvalues are specified for the $p$ bands with − (red circles) and the Ce $t_{2g}$ band with + (blue circles). For CeAs and CeSb, the valence band maximum (VBM) and conduction band minimum (CBM) at X are the pnictogen $p$ states and the Ce $t_{2g}$ states [Figs. 3(d1) and (d2)], and the $Z_2$ invariant is trivial. Since the VBM and CBM are inverted at X for CeBi [Fig. 3(d3)], the $Z_2$ invariant ends up to be nontrivial (supplementary-figure S3).

Based on the SX-ARPES results, one can now draw the topological phase diagram of CeX, as shown in Fig. 3(e). The topological phase transition is elucidated by the SOC in collaboration with the carrier compensation of the semimetallic structures. CeSb is a trivial but close to a phase transition state while CeBi is classified into a nontrivial phase due to the band inversion. The band inversion should give a rise to TSSs within the inverted band gap.

To test our conclusion about the topological band inversion, we investigate the surface dispersions on the cleaved (001) surface of CeBi by using VUV-ARPES. As reported in LaBi [26], we expect the two TSSs emerge at M̄ because the two non-equivalent X in the bulk Brillouin zone are projected to an M̄ in the surface Brillouin zone [Fig. 4(a)]. Our slab calculation presents two TSSs within the inverted band gap [s1 and s2 shown in Figs. 4(c)], which is similar to those in LaBi [26, 27]. As X and Γ are projected into Γ̄, the odd number of the TSSs should be appeared around Γ̄. However, the calculated TSS is fully buried inside the bulk continuum on the (001) surface, and thus unlikely detectable by ARPES.

In accordance with our slab calculation, we find s1 and s2 around M̄ in our VUV-ARPES results as shown in Figs. 4(d) and (e). In contrast to the previous report [28], the intrinsic surface dispersion within the band gap can be disentangled from the other irrelevant signals [marked by arrows in Fig. 4(d)]. The dispersion of s1 and s2 are clearly seen in the EDCs around M̄ [Fig. 4(e)]. Within the inverted band gap, s1 and s2 stay almost flat in en-ergy at ∼0.25 eV and ∼0.30 eV, and the both lose the spectral intensity with being close to the bulk continuum states. This overall shape of their dispersion shows good agreement with our slab calculation [Fig. 4(c)]. These results confirm the topological band inversion of CeBi determined by our bulk-sensitive ARPES.

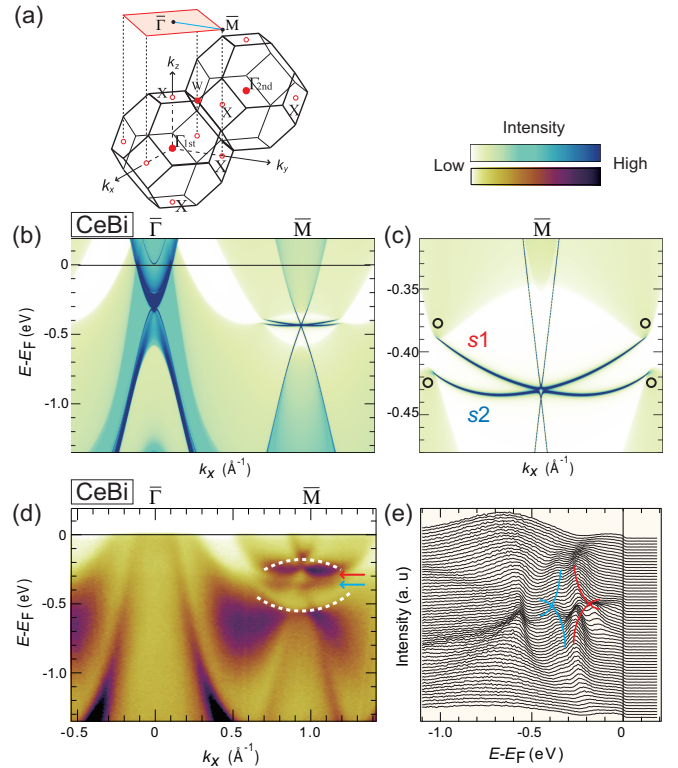

FIG. 4: (a) Bulk and (001) surface BZ of CeBi. (b) Calculated (001) surface band structures of CeBi along Γ̄-M̄ line [blue line in (a)]. For our semi-infinite slab calculations, two CeBi layers are regarded as a unit for the calculation scheme of the surface Green's function presented in Ref. [33], and so the surface spectral weight is defined for the top two CeBi layers. To fit our ARPES data, $E_F$ in the calculation results is shifted to be 0.10 eV. (c) Enlarged two surface states around M̄ point. s1 and s2 label the surface bands. The circles indicate the hybridized bulk bands. (d) VUV-ARPES band map cut along Γ̄-M̄ with $h\nu$=55 eV. The dashed lines guide the inverted bulk-band dispersions taken from Fig. 3(c3). The signals from s1 and s2 are indicated by arrows. (e) The EDCs around M̄. The observed s1 and s2 are guided by red and blue lines, respectively.

In summary, we performed bulk-sensitive SX-ARPES on CeX series materials, and determined the topological phase diagram. Our experiment unambiguously demonstrated the topological transition from a trivial to a nontrivial phase across the border between CeSb and CeBi in the presented phase diagram. The mechanism is explained by the SOC in concert with the carrier-compensated semimetallic band structures. These observations provide a new insight into the band topology in the bulk of solid states.

This work was supported by Photon and Quantum Basic Research Coordinated Development Program from the Ministry of Education, Culture, Sports, Science and Technology, Japan. The SX synchrotron radiation experiments were performed with the approval of JASRI (Proposal No. 2017A1410). We thank Diamond Light Source for access to beamline i05 (SI16161-1) that contributed to the results presented here.

---